\documentclass[preprint, aps, prc]{revtex4-1}
\usepackage{hyperref}
\usepackage{graphicx}

\begin{document}
\title{QCD sum rules for the baryon octet in nuclear matter}
\author{E. L. Kryshen}
\affiliation{Petersburg Nuclear Physics Institute, Gatchina, 188300, Russia}
\date{\today}
\begin{abstract}
The baryon self-energies are expressed in terms of the QCD condensates of the lowest dimension in symmetric and asymmetric nuclear matter within the QCD sum-rule approach. The self-energies are shown to satisfy the Gell-Mann--Okubo relations in the linear $SU(3)$ breaking approximation. The results are in qualitative agreement with those obtained by the standard nuclear physics methods.
\end{abstract}
\maketitle
\section{Introduction}
The study of the in-medium interactions of the octet of baryons is one of the hot topics in nuclear physics. While many years of theoretical and experimental investigation provided a very precise phenomenology of nucleon interactions, the hyperon properties in nuclear matter remain much less known.

The experimental information on the hyperon in-medium interactions mainly comes from the hypernuclear physics. In the past years, an impressive experimental data on $\Lambda$--hypernuclei has been accumulated providing a potential depth $U_\Lambda \approx - 30$ MeV at the saturation density $\rho_0=0.17 {\rm\ fm^{-3}}$~\cite{Millener1988}. On the other hand, the unavailing search for $\Sigma$ hypernuclear states~\cite{Bart1999} and the study of $\Sigma^-$ atoms~\cite{Batty1994,Batty1994a,Mares1995} show strong evidence for a repulsive nature of the $\Sigma$ hyperon potential in nuclear matter. The $\Xi$ nuclear interactions seem to be attractive with the potential $U_\Xi \approx -18$ MeV~\cite{Dover1983,Fukuda1998,Khaustov2000}. Finally, the hyperon-hyperon interactions were not really measured, there is just a handful of double $\Lambda$ hypernuclear events~\cite{Gal2005}. Of course, the hypernuclear data is limited to the isospin-symmetric matter at the saturation density.

However, the hyperon in-medium potentials are essential for the determination of the composition of neutron star matter~\cite{SchaffnerBielich2008,SchaffnerBielich2010}. For example, in case of an attractive $\Sigma$ hyperon potential, the $\Sigma^-$ can appear even before the $\Lambda$ hyperon in dense matter, but if $\Sigma$ hyperon potential is repulsive then $\Sigma$ hyperons are not populated at all. The hyperon population largely influences the mass-radius relation and maximum mass of neutron stars, the cooling of neutron stars, the stability with regard to the emission of gravitational waves and the possible early onset of the QCD phase transition in the neutron star cores~\cite{SchaffnerBielich2008}. Finally, the knowledge of hyperon in-medium properties is required to investigate an exciting possibility of strange hadronic matter~\cite{Schaffner-Bielich2000} stable against strong-interaction decays.

There are several theoretical approaches to the problem of hyperon interactions in nuclear matter. The traditional method is based on the relativistic mean-field approximation (RMF) with effective meson-hyperon couplings fixed from the hypernuclear data and supplemented by the flavour $SU(3)$ considerations~\cite{Glendenning1991,Schaffner1996}. A wide range of predictions was obtained in this approach, depending on the set of parameters chosen to describe the avaliable data.

Another commonly used method is based on the Brueckner-Hartree-Fock (BHF) approximation~\cite{Schulze1998, Schulze2006, Yamamoto2000} with the soft-core hyperon-nucleon potentials~\cite{Maessen1989,Stoks1999} extracted from the $YN$ scattering experiments. In general, BHF approach successfully reproduces available hypernuclear data, however, the uncertainties in the scarce hyperon scattering data lead to large differences in the resulting~$YN$ in-medium potentials.

One more method  for the calculation of the $\Lambda$ and $\Sigma$ mass shifts is provided by the chiral perturbation theory (ChPT)~\cite{Korpa2001}. However, this approach is limited to low densities up to $0.4 \rho_0$, and, even in this region, uncertainties are high since one has to determine 12 low-energy constants from the fits to the scarce experimental data on the hyperon-nucleon scattering.

Finally, there is a possibility to study the density dependence of the hyperon properties within the QCD sum rule approach which is based on the dispersion relations for the correlation functions of corresponding hadronic currents. Initially, QCD sum rule approach was developed to express vacuum characteristics of mesons in terms of expectation values of QCD operators known as ``condensates''~\cite{Shifman1979}. The sum rule approach was also applied to the calculation of nucleon properties in vacuum~\cite{Ioffe1981,Ioffe1984}. Later, it was successfully extended to the studies of nucleon self-energies in nuclear matter~\cite{Drukarev1988,Drukarev1990,Drukarev2011} as well as to the calculation of the nucleon-nucleus scattering amplitude~\cite{Henley1993} and in-medium modifications of vector meson properties~\cite{Hatsuda1992}. An alternative version of the finite-density QCD sum rules, based on the Lehmann representation for the Green function, has been also developed~\cite{Furnstahl1992} and applied to the calculation of $\Lambda$ and $\Sigma$ hyperon properties in symmetric nuclear matter~\cite{Jin1994a,Jin1995b,Zhong2007,Li2008}. In contrast to other nuclear physics methods, the finite-density sum rule approach does not rely on phenomenological baryon parameters. The baryon in-medium properties are expressed in terms of the QCD condensates, which can be either calculated in a model-independent way or related to observables.

In this paper, the finite-density QCD sum rule approach~\cite{Drukarev2011} is extended to the calculation of the baryon octet properties in nuclear matter. The baryon self-energies are expressed in terms of the lowest dimension quark and gluon condensates taken in the gas approximation. It is known that the $SU(3)$ symmetry breaking in the baryon octet in-medium properties is caused both by the nonvanishing strange quark mass $m_s$ and the $SU(3)$ asymmetric quark composition of the medium itself. We will show in this paper that the baryon self-energies satisfy the Gell-Mann--Okubo relations in the linear $SU(3)$ breaking approximation. The numerical results are obtained for the symmetric and asymmetric nuclear matter and studied as functions of the scalar quark condensate. Compared to papers~\cite{Jin1994a,Jin1995b,Zhong2007,Li2008}, we concentrate on the effects of the broken $SU(3)$ symmetry in the baryon octet and extend the hyperon sum rules to the case of asymmetric nuclear matter providing a framework for the calculation of the neutron star equation of state within the QCD sum rules approach. The properties of the $\Xi$ hyperon in nuclear matter are considered for the first time.

The structure of the paper is as follows. In Sec.~\ref{sec:vac}, the QCD sum rules for the baryon octet in vacuum are reviewed. In Sec.~\ref{sec:mat} the finite-density QCD sum rules are generalized to the case of the matter consisting of an arbitrary mixture of baryons, the approximate solutions are obtained and the numerical results in the symmetric and asymmetric nuclear matter are discussed. In Sec.~\ref{sec:dis}, the results are compared to experimental data and predictions of other approaches. A summary is provided in Sec.~\ref{sec:sum}.

\section{Baryon octet in vacuum}
\label{sec:vac}

The QCD sum rule approach in vacuum is based on the dispersion relation for the correlation function
\begin{equation}
 \Pi_{B0}(q^2) = i \int d^4 x e^{i q\cdot x} \langle 0 |\,{\cal T}\, j_B(x) \bar j_B(0)\,| 0 \rangle,
\end{equation}
where ${\cal T}$ denotes the time-ordered product and $j_B$ is the local three-quark current with quantum numbers of the baryon in interest. The usual choice for the proton current is~\cite{Ioffe1981,Ioffe1983}:
\begin{equation}
 j_p = \epsilon_{abc} (u_a^T C \gamma_\mu u_b)\gamma_5 \gamma^\mu d_c\,,
\end{equation}
where $a$, $b$ and $c$ are the color indices, $T$ denotes the transpose and $C$ is the the charge conjugation matrix. The currents for other members of the baryon octet may be obtained by the flavour $SU(3)$ transformation of the proton current~\cite{Reinders1985}.

The general idea of the QCD sum rules is to approach the bound state problem in QCD from the asymptotic freedom side. At large negative $q^2$, the correlation function  $\Pi_{B0}(q^2)$ is approximated by the power series in $q^{-2}$ known as the Operator Product Expansion (OPE)~\cite{Wilson1969}. On the other hand, the imaginary part of $\Pi_{B0}(q^2)$ at $q^2>0$ can be described in terms of the observable hadrons. Two momentum regions are connected in the dispersion relation for the function $\Pi_{B0}(q^2)$:
\begin{equation}
 \Pi_{B0}(q^2) = \frac{1}{\pi} \int \frac{{\rm Im\,} \Pi_{B0}(k^2)}{k^2-q^2}\,dk^2.
\label{eq:DR0}
\end{equation}
At large negtive $q^2$, the left-hand side of this equation is approximated by several lowest terms of OPE $\Pi_{B0}(q^2) \approx \Pi_{B0}^{\rm OPE}(q^2)$ with the coefficients containing the expectation values of the local quark and gluon field operators.

The phenomenological right-hand side of equation~(\ref{eq:DR0}) is usually considered in the ``pole $+$ continuum'' model where the lowest lying pole in ${\rm Im\,} \Pi_{B0}(k^2)$, corresponding to the baryon in interest, is separated from the higher $k^2$ singularities approximated by the continuum~\cite{Shifman1979}:
\begin{equation}
 {\rm Im\,} \Pi_{B0}(k^2) = \lambda_B^2 \delta(k^2 -m_B^2) +\frac{1}{2i} \theta(k^2 - W_B^2)\Delta \Pi^{\rm OPE}_{B0}(k^2)
\end{equation}
where $\lambda_B$ and $m_B$ are the baryon residue and mass while $W_B$ represents the effective continuum threshold value. $\Delta \Pi^{\rm OPE}_{B0}(k^2)$ denotes a discontinuity caused by logarithmic terms of the perturbative expansion of the correlation function. Thus the dispersion relation~(\ref{eq:DR0}) takes the form:
\begin{equation}
 \Pi_{B0}^{\rm OPE}(q^2) =
 \frac{\lambda_B^2}{m_B^2 - q^2}
+ \frac{1}{2\pi i} \int_{W_B^2}^\infty \frac{\Delta \Pi_{B0}^{\rm OPE}(k^2)}{k^2-q^2} dk^2.
\label{eq:DR0approx}
\end{equation}

The perturbative expansion at the left-hand side becomes increasingly valid for large $|q^2|=Q^2$, while the ``pole $+$ continuum'' assumption becomes more accurate when $|q^2|$ decreases. Usually, a certain intermediate region of $q^2$ values is considered where approximations on both sides of equation~(\ref{eq:DR0}) are believed to be valid. To improve the overlap of two approximations, the Borel transform is usually applied to both sides of equation~(\ref{eq:DR0approx}) converting a function of $Q^2$ into the function of the Borel mass $M^2$~\cite{Shifman1979}.

The correlation function $\Pi_{B0}(q^2)$ can be decomposed into two structures:
\begin{equation}
\Pi_{B0}(q^2) = \hat q \Pi_{B0}^q(q^2) + I \Pi_{B0}^I(q^2)\,,
\end{equation}
where $I$ represents the identity matrix and $\hat q = q_\mu \gamma^\mu$. The Borel transformed dispersion relations for the structures $\Pi_{B0}^q(q^2)$ and $\Pi_{B0}^I(q^2)$ are known as QCD sum rules in vacuum~\cite{Ioffe1981}:
\begin{eqnarray}
 \Pi_{B0}^{q}(M^2)- \int_{W_B^2}^\infty \frac{\Delta \Pi_{B0}^{q}(k^2)}{2\pi i} e^{-k^2/M^2} dk^2
 &=&  \lambda_B^2 e^{-m_B^2/M^2}
\label{eq:SR0_preliminary1}
\\
 \Pi_{B0}^{I}(M^2)- \int_{W_B^2}^\infty \frac{\Delta \Pi_{B0}^{I}(k^2)}{2\pi i} e^{-k^2/M^2} dk^2
 &=&  m_B\lambda_B^2 e^{-m_B^2/M^2}
\label{eq:SR0_preliminary2}
\end{eqnarray}
It is convenient to write these equations in a compact form:
\begin{equation}
{\cal L}_{B0}^l(M^2,W_B^2) = {\cal R}_{B0}^l(M^2,m_B,\tilde \lambda_B^2), \qquad l=q,I
\label{eq:SR0}
\end{equation}
where ${\cal L}_{B0}^l(M^2,W_B^2)$ represent the left-hand sides of the sum rule equations~(\ref{eq:SR0_preliminary1}--\ref{eq:SR0_preliminary2}) multiplied by the factor $32 \pi^4$ while the right-hand sides are expressed as
\begin{eqnarray}
{\cal R}_{B0}^q(M^2,m_B,\tilde \lambda_B^2) &=& \tilde \lambda_B^2 e^{-m_B^2/M^2}, \\
{\cal R}_{B0}^I(M^2,m_B,\tilde \lambda_B^2) &=& m_B \tilde \lambda_B^2 e^{-m_B^2/M^2}
\end{eqnarray}
with $\tilde \lambda_B^2 = 32 \pi^4 \lambda_B^2$.

The sum rules for baryons are usually considered in a certain interval of the Borel mass $M^2$ where the contribution of the continuum and higher OPE corrections are found to be relatively small~\cite{Belyaev1982}:
\begin{equation}
 0.8{\rm\ GeV^2} < M^2 < 1.4 {\rm\ GeV^2}.
\label{eq:M}
\end{equation}
The unknown values $m_B$, $\lambda_B$ and $W_B^2$ are obtained by minimization of the function
\begin{equation}
 \chi^2(m_B,\tilde \lambda_B^2,W_B^2) =
\sum_j \sum_{l=q,I}
\left(
\frac{{\cal L}_{B0}^l(M_j) - {\cal R}_{B0}^l(M_j)}
{{\cal L}_{B0}^l(M_j)}\right)^2\,,
\label{eq:chi2}
\end{equation}
which insures the most accurate approximation of ${\cal L}_{B0}^l(M^2,W_B^2)$ by ${\cal R}_{B0}^l(M^2,m_B,\tilde \lambda_B^2)$ with $M_j$ being a set of points evenly spaced within the fiducial interval~(\ref{eq:M}). There is also an alternative approach based on the logarithmic measure~\cite{Belyaev1982,Leinweber1990}.

Initially, the QCD sum rule approach was applied to the calculation of the baryon masses in~\cite{Ioffe1981,Ioffe1984,Belyaev1982,Belyaev1983}. In the nucleon case, the following expressions were obtained for the left-hand sides:
\begin{eqnarray}
{\cal L}_{N0}^q(M^2,W_N^2) &=& A_0+A_4 b + A_6 a^2 +A_8 \mu_0^2 a^2\,,
\label{eq:N1}
\\
{\cal L}_{N0}^I(M^2,W_N^2) &=& B_3 a + B_7 a b +B_9 a^3\,,
\label{eq:N2}
\end{eqnarray}
where $\mu_0^2 = 0.8 {\rm\ GeV^2}$, while for quark and gluon condensates the traditional notations are used:
\begin{equation}
 a = -(2\pi)^2 \langle 0| \bar q q| 0\rangle\,
 , \qquad
 b = (2\pi)^2 \left \langle 0\left|\frac{\alpha_s}{\pi} G^2 \right|0\right\rangle.
\end{equation}
The numerical value for the quark condensate $\langle 0| \bar q q| 0\rangle = -(0.24 {\rm\;GeV})^3$ was obtained from the well-known Gell-Mann--Oakes--Renner relation~\cite{Gell-Mann1968} while the value of the gluon condensate $\left \langle 0\left|\frac{\alpha_s}{\pi} G^2 \right|0\right\rangle = (0.33 {\rm\;GeV})^4$ was extracted from the analysis of leptonic decays of $\rho$ and $\phi$ mesons and supported by the QCD sum rule analysis of the charmonium spectrum~\cite{Vainshtein1978}. Note that isotopic invariance is assumed for the light quark condensates with $q$ denoting the light quark field, while higher-dimensional quark condensates are considered in the factorization approximation: $ \langle 0 |\bar q q \bar q q |0\rangle = (\langle 0|\bar q q |0\rangle)^2$~\cite{Shifman1979}.

The coefficients $A_n$ and $B_n$ are functions of $M^2$ and $W_N^2$ with the subscript denoting the dimension of the corresponding condensate:
\begin{eqnarray}
A_0 &=& \frac{M^6 E_2}{L^{4/9}},\qquad
A_4 = \frac{M^2 E_0}{4L^{4/9}} ,\qquad
A_6 = \frac{4}{3}L^{4/9},\qquad
A_8 = -\frac{1}{3M^2L^{2/27}}, \nonumber\\
B_3 &=& 2 M^4 E_1,\qquad
B_7 = -\frac{1}{9},\qquad
B_9 = \frac{272\alpha_s}{81\pi M^2 L^{1/9}},\nonumber
\label{eq:AB}
\end{eqnarray}
These expressions depend on the continuum threshold via the functions $E_n$ defined as:
\begin{equation}
E_n =1 - e^{-x} \sum_{k=0}^n \frac{x^k}{k!}, \qquad x = \frac{W_B^2}{M^2}
\end{equation}
Finally, the function $L=L(M^2)$ accounts for the leading logarithmic corrections~\cite{Shifman1979}:
\begin{equation}
 L(M^2) = \frac{\ln M^2/\Lambda^2}{\ln \nu^2/\Lambda^2}
\end{equation}
Here $\Lambda=\Lambda_{\rm{QCD}} = 0.15$ GeV, while $\nu=0.5$ GeV is the OPE normalization point.

The expression for the nucleon mass as function of Borel mass $M$ and continuum threshold $W_N$ directly follows from~(\ref{eq:SR0}), (\ref{eq:N1}) and (\ref{eq:N2}):
\begin{equation}
m_N(M^2,W_N^2) =
\frac{B_3 a +B_7 ab +B_9 a^3}{A_0+A_4 b +A_6 a^2 +A_8 \mu_0^2 a^2}
\label{eq:mN}
\end{equation}
Note, that this expression was obtained in the chiral $SU(2)$ limit ($m_u=m_d=0$ and $\langle 0|\bar u u |0\rangle = \langle 0|\bar d d |0\rangle$). In the case of hyperons, the strange quark mass $m_s$ and the difference in values of strange an light quark condensates become important. The value of the strange quark mass $m_s$ is about $150$ MeV with the uncertainty of $20\%$~\cite{Ioffe2010}. The deviation of the strange quark condensate from the light quark condensate is usually described by the parameter~$\gamma$:
\begin{equation}
 \gamma = \frac{\langle  0| \bar s s |0\rangle}{\langle 0|\bar u u |0\rangle}-1.
\end{equation}
The value $\gamma=-0.2$ is usually accepted~\cite{Belyaev1983,Reinders1985,Leinweber1990}.
Following general QCD sum rule technics, one may obtain the mass formulas for hyperons with the condensates accounted up to dimension 9~\cite{Belyaev1983,Espriu1983,Hwang1994}:
\begin{eqnarray}
m_\Lambda &=&
\frac{(B_3 +B_7b)\left(1-\frac{\gamma}{3}\right)a+B_9(1+\gamma)a^3
-\frac{1}{3}\left[S_0-S_4b-4S_6(1-\frac{\gamma}{2})a^2\right]m_s}
{A_0+A_4 b +(A_6+A_8 \mu_0^2)\left(1+\frac{4\gamma}{3}\right)a^2
+\left[\frac{1}{3} S_3(1-3\gamma)-S_5(1-\gamma)\mu_0^2\right]a m_s}
\label{eq:mL}\\
m_\Sigma &=&
\frac{(B_3 +B_7 b)(1 + \gamma)a +B_9(1 + \gamma)a^3+\left(S_0-S_4b+2 S_6a^2\right)m_s}
{A_0+A_4 b +(A_6+A_8 \mu_0^2)a^2-(S_3+S_5\mu_0^2)(1+\gamma)a m_s}
\label{eq:mS}
\\
m_\Xi &=&
\frac{(B_3+B_7 b)a+B_9(1+\gamma)^2a^3+3 S_6 (1+\gamma)a^2 m_s}
{A_0+A_4b+(A_6+A_8\mu_0^2)(1+\gamma)^2a^2-2 S_5 (1+\gamma)\mu_0^2 a m_s}
\label{eq:mX}
\end{eqnarray}
where $S_i$ are coefficients in the additional terms linear in $m_s$:
\begin{equation}
S_0 = \frac{2 M^6 E_2}{L^{8/9}}, \qquad
S_4 = \frac{M^2 E_0}{4 L^{8/9}},\qquad
S_6 = \frac{4}{3},\qquad
S_3 = \frac{2 M^2 E_0}{L^{4/9}},\qquad
S_5 = \frac{1}{3L^{26/27}}.
\end{equation}
Considering the chiral $SU(3)$ limit ($\gamma, m_s \to 0$), one can check that the expressions for hyperon masses~(\ref{eq:mL}--\ref{eq:mX}) reduce to the the nucleon mass formula~(\ref{eq:mN}). Moreover, Gell-Mann--Okubo mass relation
\begin{equation}
2(m_N + m_\Xi) = m_\Sigma + 3m_\Lambda,
\end{equation}
which is a direct consequence of the symmetry breaking in the $(\bar 3,3)$ term of the $SU(3)$ Hamiltonian,
is automatically satisfied up to the terms linear in $m_s$ and $\gamma$~\cite{Belyaev1983}. However, one has to keep in mind that this property is valid only in the limit of equal continuum threshold values $W_N^2=W_\Lambda^2=W_\Sigma^2=W_\Xi^2$.

The parameters $m_B$, $\tilde \lambda_B^2$ and $W_B^2$, which minimize the function~(\ref{eq:chi2}), are shown in Table~\ref{tab:fit}. The obtained baryon masses agree with the experimental values within 10\% accuracy. The systematic underestimates can be attributed to the fact that radiative corrections~\cite{Sadovnikova2005} or non-perturbative effects due to instantons~\cite{Dorokhov1990,Forkel1993} may be important.

\begin{table}[htb]
 \begin{tabular}{|c|c|c|c|c|c|}
\hline
$B$       & $m_B^{\rm exp}$, GeV & $m_B$, GeV& $\tilde \lambda_B^2$, GeV$^6$ & $W_B^2$, GeV$^2$  \\
\hline
$N$       & 0.940 & 0.934 & 1.897 & 2.119 \\
$\Lambda$ & 1.116 & 1.103 & 3.189 & 3.069 \\
$\Sigma$  & 1.193 & 1.104 & 3.066 & 3.157 \\
$\Xi$     & 1.314 & 1.207 & 4.069 & 3.729 \\
\hline
\end{tabular}
\caption{Values of $m_B$, $\tilde \lambda_B^2$ and $W_B^2$ from the minimization procedure.}
\label{tab:fit}
\end{table}

Variation of the strange quark mass and the parameter $\gamma$ results in significant changes of the hyperon masses and their relative order. Therefore, it is instructive to consider approximate expressions for the hyperon masses as functions of the involved parameters.
Multiplying both sides of the QCD sum rules by $\exp(m_B^2/M^2)$ one may notice that the functions ${\cal L}_{B0}^l(M^2,W_B^2) e^{m_B^2/M^2}$ should not depend on the Borel mass. A simple check ensures that the leading OPE terms, multiplied by the factor $\exp(m_B^2/M^2)$, can also be approximated by the constants in the range~(\ref{eq:M}) within 10\% accuracy. Referring to this observation, let us introduce the notation:
\begin{equation}
 \bar X_n(m_B,W_B^2,\tilde \lambda_B^2) = \frac{\overline{X_n(M^2,W_B^2)\exp(m_B^2/M^2)}}{\tilde \lambda_B^2},
\label{eq:Xn}
\end{equation}
where $X_n$ stands for the functions $A_n$, $B_n$ or $S_n$, while the overline denotes averaging over the Borel mass range~(\ref{eq:M}). The numerical values for $\bar A_n$, $\bar B_n$ and $\bar S_n$ were calculated at values of $m_B$, $W_B^2$ and $\tilde \lambda_B^2$ from the minimization procedure and provided in Table~\ref{tab:Xn}. Then the mass formulas~(\ref{eq:mN}), (\ref{eq:mL}), (\ref{eq:mS}), (\ref{eq:mX}) can be expressed in terms of the averaged values $\bar A_n$, $\bar B_n$ and $\bar S_n$ instead of $M^2$--dependent functions. Such expressions reproduce the values of the baryon masses, fitted via $\chi^2$~(\ref{eq:chi2}), with an accuracy of the order of $0.5\%$. For example, in the case of nucleon and $\Xi$ hyperon, one can write:
\begin{eqnarray}
\bar{m}_N &=&
\frac{1.61a-0.10ab+0.43a^3}{0.38+0.22 b+ 1.97a^2-0.37\mu_0^2 a^2 },\\
\bar{m}_\Xi &=&
\frac{1.91a-0.08ab +0.36(1+\gamma)^2a^3+3.93 (1+\gamma)a^2 m_s}
{0.65+0.20 b+(1.61-0.31\mu_0^2)(1+\gamma)^2a^2- 0.42(1+\gamma)\mu_0^2 a m_s},
\end{eqnarray}
where all values are expressed in powers of GeV. To our knowledge, such expressions for the baryon masses are considered for the first time.
They reveal relative contributions of different OPE terms providing a convenient way to study the dependence on the condensate values. Similarly, using the expressions for the hyperon masses, it is easy to show that reasonable variations of $m_s$ and $\gamma$ values do not allow tuning all hyperon masses to their experimental values. Therefore, we will apply conventional values of $m_s=150$ MeV and $\gamma = -0.2$ in the finite-density QCD sum rule analysis.

\begin{table}[htb]
\label{tab:Xn}
\begin{tabular}{|c|c|c|c|c|c|c|c|c|c|c|c|c|}
\hline
$B$      & $\bar A_0$ & $\bar A_4$ & $\bar A_6$ & $\bar A_8$ & $\bar B_3$ & $\bar B_7$ & $\bar B_9$ & $\bar S_0$ & $\bar S_3$ & $\bar S_4$ & $\bar S_5$ & $\bar S_6$\\
\hline
$N$     &$0.38$ &$0.22$ &$1.97$ &$-0.37$ &$1.61$ &$-0.10$ &$0.43$ & $-$ & $-$ & $-$ & $-$ & $-$\\
$\Lambda$&$0.54$ &$0.20$ &$1.63$ &$-0.31$ &$1.76$ &$-0.08$ &$0.36$ &$0.87$ &$1.61$ &$0.16$ &$0.21$ &$1.32$ \\
$\Sigma$ &$0.58$ &$0.21$ &$1.70$ &$-0.32$ &$1.87$ &$-0.09$ &$0.37$ &$0.94$ &$1.69$ &$0.17$ &$0.22$ &$1.38$ \\
$\Xi$    &$0.65$ &$0.20$ &$1.61$ &$-0.31$ &$1.91$ &$-0.08$ &$0.36$ &$-$ &$-$ &$-$ &$0.21$ &$1.31$ \\
\hline
\end{tabular}
\caption{Values $\bar A_n$, $\bar B_n$ and $\bar S_n$ averaged over the Borel mass range according to equation~(\ref{eq:Xn}) in powers of GeV.}
\end{table}

\section{Baryon octet in nuclear matter}
\label{sec:mat}
In this section, we will develop the framework for the calculation of the baryon octet parameters in nuclear matter following the finite-density sum rule approach reviewed in~\cite{Drukarev2011}.

\subsection{QCD sum rules in nuclear matter}
The propagation of a system with four-momentum $q$ in nuclear matter is described by the correlation function:
\begin{equation}
 \Pi_{Bm}(q^2) = i \int d^4 x e^{i q\cdot x} \langle M |\,{\cal T}\, j_B(x) \bar j_B(0)\,| M \rangle,
\end{equation}
where $|M \rangle$ is the ground state of nuclear matter. Considering the nuclear matter as a system of $A$ nucleons with momenta $p_i$, one may introduce the vector
\begin{equation}
 p = \frac{\sum p_i}{A}
\end{equation}
which turns to $p\approx (m,0)$ in the rest frame of the matter, where $m$ is the nucleon mass. The spectrum of the function $\Pi_{Bm}(q^2)$ appears to be much more complicated than that of $\Pi_{B0}(q^2)$, however by fixing the value of $s = (p+q)^2$, one may separate the singularities connected with the matter itself from those connected with the baryon in the matter~\cite{Drukarev1990,Drukarev2011,Henley1993}. In this paper, the effects of the nucleon Fermi motion on the baryon properties will be neglected, therefore the threshold value of $s=(m+m_B)^2$ will be used in the calculations.

The general form of the polarization function in the nuclear matter can be presented as:
\begin{equation}
 \Pi_{Bm}(q) =\hat q \Pi_{Bm}^q(q^2,s) + I \Pi_{Bm}^I(q^2,s)+ \frac{\hat p}{m} \Pi_{Bm}^p(q^2,s).
\end{equation}
The in-medium QCD sum rules are then derived as the Borel-transformed dispersion relations for the components $\Pi_{Bm}^i(q^2,s)$:
\begin{equation}
 \Pi_{Bm}^i(q^2,s) = \frac{1}{\pi} \int \frac{{\rm Im\;} \Pi_{Bm}^i(k^2,s)}{k^2-q^2} dk^2,
\quad i = q, I, p.
\end{equation}
It was shown that the spectrum of the function $\Pi_{Bm}(q^2,s)$ can be described by the ``pole+continuum'' model similar to the vacuum case at least until the terms of the order $\rho^2$ are included in the OPE~\cite{Drukarev2011}. One may consider a general expression for the propagator of the baryon $B$ in nuclear matter:
\begin{equation}
 G_B^{-1} = (G_B^0)^{-1} - \Sigma_B,
\end{equation}
where $G_B^0 = (\hat q - m_B)^{-1}$ is the free baryon propagator and $\Sigma_B$ is a general expression for the baryon self-energy in nuclear matter:
\begin{equation}
 \Sigma_B = \hat q \Sigma_B^q + \frac{\hat p}{m} \Sigma_B^p + I \Sigma_B^I.
\end{equation}
Inverting $G_B^{-1}$, we find for the in-medium baryon propagator:
\begin{equation}
G_B = Z_B \frac{\hat q -\hat p \Sigma_B^V/m + m_B^*}{q^2-m_{Bm}^2},
\label{eq:GB}
\end{equation}
where $Z_B = [(1-\Sigma_q)(1+\Sigma_B^V/m)]^{-1}$, while $\Sigma_B^V$ and $m_B^*$ correspond to the vector self-energy and the effective mass in nuclear physics:
\begin{equation}
 \Sigma_B^V = \frac{\Sigma^P_B}{1-\Sigma_B^q}, \qquad m_B^* = \frac{m_B + \Sigma^I_B}{1 -\Sigma_B^q}.
\end{equation}
For the new position of the baryon pole $m_{Bm}$ we find:
\begin{equation}
 m_{Bm}^2 = \frac{(s-m^2)\Sigma_B^V/m - (\Sigma_B^V)^2 + m_B^{*2}}{1+ \Sigma_B^V/m}.
\end{equation}
Following definitions, accepted in nuclear physics, it is also convenient to introduce the scalar self-energy $\Sigma_B^S = m_B^* - m_B$ and the non-relativistic baryon potential $U_B=\Sigma_B^V+\Sigma_B^S$.

The Borel transformed sum rule equations take the form ($l=q,I,p$):
\begin{equation}
{\cal L}_{Bm}^l(M^2,W_{Bm}^2) = {\cal R}_{Bm}^l(M^2)
\label{eq:SRm}
\end{equation}
with the phenomenological right-hand side:
\begin{equation}
{\cal R}_{Bm}^l(M^2)
= \xi^l \tilde \lambda^2_{Bm} e^{-m_{Bm}^2/M^2},
\end{equation}
where $\tilde \lambda_{Bm}^2 = 32\pi^4 Z_B \lambda_{Bm}^2$ is the effective value of the residue for the baryon $B$ in nuclear matter. The values of $\xi^l$ are determined from equation~(\ref{eq:GB}):
\begin{equation}
 \xi^q = 1,\qquad \xi^p = -\Sigma_B^V, \qquad \xi^I = m_B^*.
\end{equation}
The left-hand sides of the sum rule equations~(\ref{eq:SRm}) are calculated in the OPE approach:
\begin{equation}
{\cal L}_{Bm}^l(M^2,W_{Bm}^2) = 32 \pi^4
\left(\Pi_{Bm}^{l}(M^2,s)
- \int_{W_{Bm}^2}^\infty \frac{\Delta_{k^2} \Pi_{Bm}^{l}(k^2,s)}{2\pi i} e^{-k^2/M^2} dk^2 \right).
\label{eq:left_m}
\end{equation}
Similar to the vacuum case, we can express baryon effective masses and vector self-energies via the left-hand sides of the sum rule equations:
\begin{eqnarray}
 m^*_B(M^2,W_{Bm}^2) &=& \frac{{\cal L}_{Bm}^I(M^2,W_{Bm}^2)}{{\cal L}_{Bm}^q(M^2,W_{Bm}^2)},
\label{eq:ms}
\\
 \Sigma^V_B(M^2,W_{Bm}^2) &=& -\frac{{\cal L}_{Bm}^p(M^2,W_{Bm}^2)}{{\cal L}_{Bm}^q(M^2,W_{Bm}^2)}.
\label{eq:sv}
\end{eqnarray}

For the calculation of baryon in-medium properties, we will consider only leading OPE terms which density dependence is briefly reviewed in the next subsection.

\subsection{Condensates in nuclear matter}
\label{sec:condensates}
In this subsection, we consider the condensates in the nuclear matter of density $\rho$ consisting of an arbitrary mixture of the baryon octet members $B = p,n,\Lambda,\Sigma^{+},\Sigma^{0},\Sigma^{-},\Xi^{0},\Xi^{-}$ with concentrations $c_B$.

The lowest order of OPE in medium can be presented in terms of the vector and scalar quark condensates. The vector quark condensate is defined as:
\begin{equation}
v_\mu^i(\rho) \equiv \langle M |\bar q_i \gamma_\mu q_i| M \rangle,
\end{equation}
where $q_i$ stands for $u$, $d$ or $s$ quarks. In the rest frame of nuclear matter,
the vector condensates take the form $v_\mu^i(\rho) = v_0^i(\rho) \delta_{\mu0}$, where the functions $v^i_0(\rho)$ are linear in the nuclear matter density $\rho$:
\begin{equation}
v_0^i(\rho) = v_i \rho, \qquad
v_i =\sum_B n^i_B c_B
\end{equation}
with $n^i_B = \langle B |\bar q_i \gamma_0 q_i| B \rangle$ denoting the number of valence quarks of flavour $i$ in baryon $B$. For the ordinary nuclear matter consisting of protons and neutrons only, it is convenient to define isospin symmetric and  asymmetric combinations~\cite{Drukarev2004a}:
\begin{eqnarray}
v^{\ \ }   &=& v_u + v_d = \langle p |\bar u \gamma_0 u+\bar d \gamma_0 d| p \rangle = 3,
\label{eq:vp}\\
v^- &=& v_u - v_d = \langle p |\bar u \gamma_0 u-\bar d \gamma_0 d| p \rangle = 1.
\label{eq:vm}
\end{eqnarray}
While due to the vector current conservation the vector condensates are exactly linear in $\rho$, the scalar quark condensates $\kappa_m^i(\rho) \equiv \langle M |\bar q_i q_i| M \rangle$ are more complicated functions of density. However, in the gas approximation, they can also be expressed by the linear functions of~$\rho$:
\begin{equation}
\kappa_m^i(\rho) \approx \kappa^i_0 + \kappa_i \rho, \qquad \kappa_i = \sum_B \kappa^i_B c_B,
\end{equation}
where $\kappa^i_0 = \langle 0 |\bar q_i q_i| 0 \rangle$ while
$\kappa^i_B$ denote baryon matrix elements:
\begin{equation}
\kappa^i_B = \langle B |\bar q_i q_i| B \rangle.
\end{equation}
Following~\cite{Drukarev2004}, we introduce isospin symmetric and asymmetric combinations of the light quark expectation values:
\begin{eqnarray}
\kappa &=& \kappa^u_p + \kappa^d_p = \langle p |\bar u u + \bar d d| p \rangle,
\label{eq:kp}\\
\zeta   &=& \kappa^u_p - \kappa^d_p = \langle p |\bar u u - \bar d d| p \rangle.
\label{eq:km}
\end{eqnarray}

The expectation value $\kappa$ is directly related to the pion-nucleon sigma term $\sigma_{\pi N}$~\cite{Gasser1981}:
\begin{equation}
\kappa = \frac{2 \sigma_{\pi N}}{m_u+m_d}
\end{equation}
with $m_{u}\approx 4$ MeV and $m_{d} \approx 7$ MeV denoting the current masses of the light quarks. The $\sigma$ term can be extracted in several ways, i. e. from the subthreshold extrapolation of the $\pi N$ scattering amplitude, however there is a large discrepancy between the results~(see~\cite{Drukarev2011} for references). Assuming the conventional value of $\sigma \approx 45$ MeV~\cite{Gasser1991}, one would obtain $\kappa \approx 8$, while with the latest results of $\sigma \approx 60$ MeV, the value of $\kappa \approx 11$ is preferred~\cite{Drukarev2011}.

In contrast, the expectation value of $\zeta$ is not restricted by experimental data, therefore some model assumptions on the quark structure of the nucleon are required. If the nucleon is treated as a system of valence quarks and isospin-symmetric sea of quark-antiquark pairs, the expectation value $\zeta$ is determined by the contribution of the valence quarks. We will use the value of $\zeta=0.54$ obtained in the perturbative chiral quark model (PCQM)~\cite{Lyubovitskij2001} which was used in~\cite{Drukarev2004a} for the calculation of in-medium four-quark condensates.

The strange quark expectation value  $\kappa_p^s = \langle p |\bar s s| p \rangle$ is usually parameterized in terms of the strange quark content $y$:
\begin{equation}
 y = \frac{2\langle p | \bar s s | p \rangle}{\langle p |\bar u u + \bar d d | p \rangle}
   = \frac{2 \kappa_p^s}{\kappa}.
\label{eq:y}
\end{equation}
The parameter $y$ is strongly correlated with the value of the $\sigma$ term. The value $\sigma \approx 60$ MeV corresponds to the large $y \approx 0.35$, while the conventional value of $\sigma \approx 45$ MeV is consistent with the smaller strange quark content $y \approx 0.2$. In PCQM approach~\cite{Lyubovitskij2001a}, one gets $y = 0.08$ in support of the smaller strange quark content. We will use $y=0.08$ as a default value in this paper since the values $y=0.2$ or $0.35$ would correspond to too many strange quarks in the nucleon in contradiction with the naive non-relativistic quark model. We will also study the sensitivity of the baryon self-energies with respect to the strange quark content parameter.

The scalar quark condensate was also considered beyond the gas approximation in the framework of the meson-exchange model of nucleon-nucleon interactions~\cite{Drukarev1988,Drukarev1990}. It was shown that the nonlinear contribution to the scalar condensate may be responsible for the saturation mechanism. However, the nonlinear terms appear to be small compared to the linear term up to the saturation density, therefore they will be ignored in the framework of this paper.

The gluon condensate in nuclear matter can be also considered in the gas approximation:
\begin{equation}
 g_m \equiv \left \langle M \left| \frac{\alpha_s}{\pi} G^2\right| M  \right \rangle \approx
  g_0 + g \rho
\end{equation}
where $g_0=\left \langle 0 \left| \frac{\alpha_s}{\pi} G^2\right| 0  \right \rangle$ is the vacuum expectation value while $g$ is the nucleon matrix element:
\begin{equation}
g = \left \langle N \left| \frac{\alpha_s}{\pi} G^2\right| N  \right \rangle
\end{equation}
The value of $g$ was calculated in~\cite{Shifman1978} by averaging the trace of the QCD energy-momentum tensor. In the chiral $SU(3)$ limit, one gets $g = -\frac{8}{9}m$ which is sufficient for our analysis referring to the small contribution of the gluon condensate~\cite{Drukarev2011}.

As for the scalar quark expectation values $\langle H|\bar q_i q_i | H \rangle$ in hyperons $H=\Lambda,\Xi,\Sigma$, they cannot be directly related to observables and some model assumptions are necessary. One option is to apply Hellmann-Feynman theorem to the QCD Hamiltonian density and relate the scalar quark expectation value to the derivative $dm_H/dm_{q_i}$~\cite{Cohen1992}:
\begin{equation}
m_{q_i} \langle H| \hat q q|H \rangle = m_{q_i} \frac{d m_H}{d m_{q_i}}
\end{equation}
The functions $m_H(m_{q_i})$ and corresponding derivatives can be deduced from ChPT~\cite{Zhong2007} or vacuum QCD sum rules. Note, however, that numerical results of this paper will be limited to the case of the nonstrange nuclear matter consisting of protons and neutrons only, thus we will not need the values $\langle H| \hat q q|H \rangle$ in the present calculations. Nevertheless, our approach can be easily extended to the case of an arbitrary mixture of baryons.

\subsection{Sum rules in the gas approximation}
Following~\cite{Drukarev2011}, we can express the left-hand sides~(\ref{eq:left_m}) as a sum of vacuum expressions and terms linear in density~$\rho$:
\begin{equation}
 {\cal L}_{Bm}^{l}(M^2,W_{Bm}^2) =
 {\cal L}_{B0}^{l}(M^2,W_{Bm}^2) + X^l_B(M^2,W_{Bm}^2)\rho,
\label{eq:LBmi}
\end{equation}
where $l$ denotes the structures $q$, $I$ and $p$. Note, that ${\cal L}_{B0}^{p}(M^2,W_{Bm}^2) \equiv 0$ while the vacuum expressions ${\cal L}_{B0}^{q,I}$ are calculated at density-dependent continuum thresholds $W_{Bm}$. The functions $X^l_B$ can be expressed in terms of quark and gluon expectation values $v_i$, $\kappa_i$ and $g$, considered in the previous subsection:
\begin{eqnarray}
X^q_B &=& A^g_B g +  A^v_B \sum_i a^i_{vB} v_i + m_s A^\kappa_B \sum_i a^i_{\kappa B}  \kappa_i, 
\qquad 
\label{eq:XqBgeneral}
\\
X^I_B &=& B^\kappa_B \sum_i b^i_{\kappa B}  \kappa_i + m_s B^v_B \sum_i b^i_{vB} v_i, 
\label{eq:XIBgeneral}
\\
X^p_B &=& P^v_B \sum_i p^i_{vB} v_i,
\label{eq:XpBgeneral}
\end{eqnarray}
where the Borel transformed OPE coefficients $A^g_B$, $A^\kappa_B$, $A^v_B$, $B^\kappa_B$, $B^v_B$ and $P^v_B$ read:
\begin{eqnarray}
A^g_B &=& \frac{\pi^2 M^2 E_0}{L^{4/9}}, \quad
A^v_B = -\frac{8\pi^2}{3}\frac{(s-m^2) M^2 E_0 - M^4 E_1}{mL^{4/9}}, 
\label{eq:ABP1}
\\
A^\kappa_B &=& 4\pi^2 M^2 E_0,\quad
B^v_B = 4\pi^2\frac{(s-m^2) M^2 E_0 - M^4 E_1}{m L^{8/9}},
\\
B^\kappa_B &=& -4\pi^2 M^4 E_1, \quad
P^v_B = -\frac{32\pi^2}{3} \frac{M^4 E_1}{L^{4/9}},
\label{eq:ABP}
\end{eqnarray}
These expressions depend on the Borel mass $M^2$ and on $W_{Bm}^2$ via the functions $E_0(W_{Bm}^2/M^2)$ and $E_1(W_{Bm}^2/M^2)$ which are taken at density-dependent continuum threshold values. Note that the coefficients~(\ref{eq:ABP1}--\ref{eq:ABP}) also depend on $s$ which was fixed in the dispersion relation for the correlation function. Analogous expressions were obtained in the finite-density sum rule approach based on the Lehmann representation for the Green function~\cite{Jin1994a,Jin1995b} where the dispersion relation was written in the $q_0$ complex plane at fixed three-momentum $\bf q$. In the latter case, the Borel transformed OPE coefficients~(\ref{eq:ABP1}--\ref{eq:ABP}) appear to depend on the momentum $\bf q$ instead of the $s$ invariant.

The coefficients $a_{vB}^i$, $a_{\kappa B}^i$, $b_{vB}^i$, $b_{\kappa B}^i$ and $p_{vB}^i$ are shown in Table~\ref{tab:abp}. They depend on the baryon isospin projection $I_{3B}$ responsible for the splitting of baryon self-energies in the $(n,p)$ and $(\Xi^-,\Xi^0)$ isospin doublets and the $(\Sigma^-,\Sigma^0,\Sigma^+)$ triplet. However, in the isospin symmetric nuclear matter, the terms, proportional to $I_{3B}$, cancel out, and the isospin symmetry for the baryon self-energies is restored.

\begin{table}[htb]
\begin{tabular}{| c | c | c | c | c|}
\hline
$\hspace{0.03\textwidth}B\hspace{0.03\textwidth}$&
$\hspace{0.03\textwidth}N\hspace{0.03\textwidth}$&
$\hspace{0.03\textwidth}\Lambda\hspace{0.03\textwidth}$&
$\hspace{0.03\textwidth}\Sigma\hspace{0.03\textwidth}$&
$\hspace{0.03\textwidth}\Xi\hspace{0.03\textwidth}$ \\
\hline
$a^u_{vB}$      & 1        & $\frac56$  & $\frac{1}{2}(1+I_{3\Sigma})$ & $\frac{1}{2}(1+2I_{3\Xi})$ \\
$a^d_{vB}$      & 1        & $\frac56$  & $\frac{1}{2}(1-I_{3\Sigma})$ & $\frac{1}{2}(1-2I_{3\Xi})$ \\
$a^s_{vB}$      & 0        & $\frac26$  & $1$                  & $1$                   \\
\hline
$a^u_{\kappa B}$& 0        & $-\frac43$ & $0$                  & $0$                   \\
$a^d_{\kappa B}$& 0        & $-\frac43$ & $0$                  & $0$                   \\
$a^s_{\kappa B}$& 0        & $2$        & $1$                  & $0$                   \\
\hline
$b^u_{vB}$      & 0        & $\frac13$  & $-1-I_{3\Sigma}$             & $0$                   \\
$b^d_{vB}$      & 0        & $\frac13$  & $-1+I_{3\Sigma}$             & $0$                   \\
$b^s_{vB}$      & 0        & $-\frac23$ & $2$                  & $0$                   \\
\hline
$b^u_{\kappa B}$& $1-2I_{3N}$ & $\frac43$  & $0$                  & $1+2I_{3\Xi}$              \\
$b^d_{\kappa B}$& $1+2I_{3N}$ & $\frac43$  & $0$                  & $1-2I_{3\Xi}$              \\
$b^s_{\kappa B}$& 0        & $-\frac23$ & $2$                  & $0$                   \\
\hline
$p^u_{vB}$& $1+\frac{3}{2}I_{3N}$ & $\frac{11}{24}$ & $\frac{7}{8}(1+I_{3\Sigma})$ &$\frac{1}{8}(1+2 I_{3\Xi})$ \\
$p^d_{vB}$& $1-\frac{3}{2}I_{3N}$ & $\frac{11}{24}$ & $\frac{7}{8}(1-I_{3\Sigma})$ &$\frac{1}{8}(1-2 I_{3\Xi})$\\
$p^s_{vB}$& $0$                & $\frac{26}{24}$ & $\frac{2}{8}$        & $\frac{14}{8}$ \\
\hline
\end{tabular}
\caption{The coefficients $a_{vB}^i$, $a_{\kappa B}^i$,
$b_{vB}^i$, $b_{\kappa B}^i$ and $p_{vB}^i$ in the expressions~(\ref{eq:XqBgeneral}--\ref{eq:XpBgeneral}).}.
\label{tab:abp}
\end{table}
Let us study the symmetry properties of the obtained expressions. It is easy to check that in the chiral $SU(3)$ limit ($m_s \to 0$), the $SU(3)$ symmetry remains broken due to different coefficients $a_{vB}^i$, $b_{\kappa B}^i$ and $p_{vB}^i$ accompanying light and strange quark condensates for different baryon species. However, the $SU(3)$ symmetry is restored in the $SU(3)$ symmetric matter with equal scalar and vector quark compositions ($v_u=v_d=v_s$ and $\kappa_u=\kappa_d=\kappa_s$), providing degenerate functions $X^l_B$ and hence equal effective masses~(\ref{eq:ms}) and vector self-energies~(\ref{eq:sv}) for the baryon octet. Besides, in the isospin symmetric matter, the functions $X^l_B$ satisfy relations similar to the Gell-Mann--Okubo mass formula:
\begin{equation}
2(X^l_N + X^l_\Xi) = X^l_\Sigma + 3X^l_\Lambda
\end{equation}
which are valid in the limit of equal effective continuum thresholds for the terms, proportional to the gluon and light quark condensates, and for the terms linear in $m_s$, $\kappa_s$ or $v_s$. Under this approximation, the same relations also hold for the effective baryon masses and vector self-energies:
\begin{eqnarray}
2(m^*_N      + m^*_\Xi) &=& m^*_\Sigma + 3m^*_\Lambda,
\label{eq:GMOms}
\\
2(\Sigma^V_N + \Sigma^V_\Xi) &=& \Sigma^V_\Sigma + 3\Sigma^V_\Lambda.
\label{eq:GMOsv}
\end{eqnarray}
Of course, these relations should be valid for any model based on $SU(3)$ symmetry breaking hypothesis.

Exact solutions for the baryon effective masses and vector self-energies, as well as in-medium effective thresholds $W_{Bm}^2$ and residues $\lambda_{Bm}^2$, could be found by minimization of the function similar to~(\ref{eq:chi2}). Note, however, that exact solutions of the nucleon sum rules, accounting for four-quark condensates and nonlocalities of the lowest dimension condensates, resulted only in slight changes of the continuum threshold $W_{Nm}^2$ up to the saturation density~\cite{Drukarev2004a}. Therefore we can safely consider approximate solutions described in the next subsection.

\subsection{Approximate solution}
In this subsection, we consider an approximate solution for the sum rule equations~(\ref{eq:ms}) and~(\ref{eq:sv}) by
replacing the in-medium continuum threshold $W_{Bm}^2$ by its vacuum value $W_B^2$~\cite{Drukarev2011}.
\begin{eqnarray}
m^*_B(M^2,W_{B}^2) &=& \frac{{\cal L}_{B0}^I(M^2,W_{B}^2)+X^I_B(M^2,W_{B}^2) \rho}{{\cal L}_{B0}^q(M^2,W_{B}^2)+X^q_B(M^2,W_{B}^2) \rho},\qquad \\
\Sigma^V_B(M^2,W_{B}^2) &=& \frac{-X^p_B(M^2,W_{B}^2) \rho}{{\cal L}_{B0}^q(M^2,W_{B}^2)+X^q_B(M^2,W_{B}^2) \rho}.\qquad
\end{eqnarray}
We can divide both numerators and denominators by ${\cal L}_{B0}^q(M^2,W_{B}^2)$ and express $m^*_B$ and $\Sigma^V_B$ in the form which is generally accepted in the nuclear physics:
\begin{eqnarray}
m^*_B(M^2,W_{B}^2) &=& \frac{m_B(M^2,W_{B}^2)+{\cal F}^I_B(M^2,W_{B}^2)}{1+{\cal F}^q_B(M^2,W_{B}^2)},\qquad
\label{eq:msB}
 \\
\Sigma^V_B(M^2,W_{B}^2) &=& -\frac{{\cal F}^p_B(M^2,W_{B}^2)}{1+{\cal F}^q_B(M^2,W_{B}^2)}, \qquad
\label{eq:svB}
\end{eqnarray}
where
\begin{equation}
{\cal F}^l_B(M^2,W_{B}^2) = \frac{X^l_B(M^2,W_{B}^2)}{{\cal L}_{B0}^q(M^2,W_{B}^2)} \rho.
\end{equation}
Applying the vacuum sum rule, one gets:
\begin{equation}
{\cal F}^l_B(M^2,W_{B}^2,\tilde \lambda_B^2) = \frac{X^l_B(M^2,W_{B}^2)\exp(m_B^2/M^2)}{\tilde \lambda_B^2} \rho.
\end{equation}
Similar to the vacuum case, one can average the functions ${\cal F}^l_B$ over the Borel mass range at the vacuum continuum threshold and express them in terms of
the average values for the functions $\bar A_g$,
$\bar A_v$, $\bar A_\kappa$, $\bar B_v$, $\bar B_\kappa$, $\bar P_v$:
\begin{eqnarray}
\bar{\cal F}^q_B &=&\left( \bar A^g_B g + \bar  A^v_B \sum_i a^i_{vB} v_i + m_s\bar  A^\kappa_B \sum_i a^i_{\kappa B}  \kappa_i\right) \rho, \qquad \\
\bar {\cal F}^I_B &=&\left(\bar B^\kappa_B \sum_i b^i_{\kappa B}  \kappa_i + m_s\bar  B^v_B \sum_i b^i_{vB} v_i\right) \rho, \\
\bar{\cal F}^p_B &=&\left(\bar P^v_B \sum_i p^i_{vB} v_i\right)\rho,
\end{eqnarray}
where the meaning of the overline was defined in~(\ref{eq:Xn}). The values $\bar A_g$, $\bar A_v$, $\bar A_\kappa$, $\bar B_v$, $\bar B_\kappa$, $\bar P_v$, averaged at vacuum continuum thresholds for $N$, $\Lambda$, $\Sigma$ and $\Xi$ baryons, are shown in Table~\ref{tab:Abar}. Then the effective baryon masses and the vector self-energies~(\ref{eq:msB}--\ref{eq:svB}) can be expressed in terms of average values $\bar{ \cal F}^l_B$ and $\bar m_B$:
\begin{equation}
\bar m_B^* = \frac{\bar m_B+\bar {\cal F}_B^I}{1+\bar {\cal F}_B^q}, \qquad
\bar \Sigma_V = -\frac{\bar {\cal F}_B^p}{1+\bar {\cal F}_B^q}.
\label{eq:approx_ms_sv}
\end{equation}

In the next subsection, we will study these approximate solutions in the case of nonstrange nuclear matter.

\begin{table}[htb]
\begin{tabular}{| c | r | r | r | r|}
\hline
$\hspace{0.03\textwidth}B\hspace{0.03\textwidth}$&
$\hspace{0.03\textwidth}N\hspace{0.03\textwidth}$&
$\hspace{0.03\textwidth}\Lambda\hspace{0.03\textwidth}$&
$\hspace{0.03\textwidth}\Sigma\hspace{0.03\textwidth}$&
$\hspace{0.03\textwidth}\Xi\hspace{0.03\textwidth}$ \\
\hline
$\bar A^g_B     $  &$      9$ &$      8$ &$      8$ &$      8$  \\
$\bar A^v_B     $  &$    -48$ &$    -55$ &$    -65$ &$    -74$  \\
$\bar A^\kappa_B$  &$     44$ &$     39$ &$     41$ &$     40$  \\
$\bar B^\kappa_B$  &$    -32$ &$    -35$ &$    -37$ &$    -38$  \\
$\bar B^v_B     $  &$     58$ &$     68$ &$     80$ &$     90$  \\
$\bar P^v_B     $  &$    -69$ &$    -75$ &$    -80$ &$    -82$  \\
\hline
\end{tabular}
\caption{The values $\bar A_g$, $\bar A_v$, $\bar A_\kappa$, $\bar B_v$, $\bar B_\kappa$, $\bar P_v$, averaged at vacuum continuum thresholds (in powers of GeV). The numbers are round off to integer values.}
\label{tab:Abar}
\end{table}

\subsection{Symmetric nuclear matter}
Let us consider the nuclear matter, composed of protons and neutrons with baryonic concentrations $c_p$ and $c_n$ respectively. We can also define the isospin asymmetry parameter $\beta = c_n - c_p$, which is equal to 1 in the pure neutron matter.
Then the functions $X^l_B$ can be expressed in terms of the nucleon matrix elements $v$, $v^-$, $\kappa$, $\zeta$, and the strange quark content $y$:
\begin{eqnarray}
X^q_B &=& A_B^g g + A_B^v (a^{+}_{vB} v+a^{-}_{vB} v^- I_{3B} \beta)
 + m_s A_B^\kappa(a^{+}_{\kappa B} \kappa + a^{y}_{\kappa B} \kappa y),
 \label{eq:XBq}
\qquad \\
X^I_B &=& B_B^\kappa (b^{+}_{\kappa B} \kappa +b^{-}_{\kappa B}\zeta I_{3B} \beta
+ b^{y}_{\kappa B} \kappa y)
 + m_s B_B^v (b^{+}_{vB} v+b^{-}_{vB} v^- I_{3B} \beta),
 \label{eq:XBI}
 \\
X^p_B &=& P_B^v (p^{+}_{vB} v+p^{-}_{vB} v^- I_{3B} \beta)
 \label{eq:XBp}
\end{eqnarray}
with coefficients $a^\pm_{vB}$, $b^\pm_{vB}$, $a^+_{\kappa B}$, $b^\pm_{\kappa B}$, $a^{y}_{\kappa B}$, $b^{y}_{\kappa B}$ and $p^\pm_{vB}$ summarized in Table~\ref{tab:abp2}.
\begin{table}[htb]
\begin{tabular}{| c | c | c | c | c|}
\hline
$\hspace{0.03\textwidth}B\hspace{0.03\textwidth}$&
$\hspace{0.03\textwidth}N\hspace{0.03\textwidth}$&
$\hspace{0.03\textwidth}\Lambda\hspace{0.03\textwidth}$&
$\hspace{0.03\textwidth}\Sigma\hspace{0.03\textwidth}$&
$\hspace{0.03\textwidth}\Xi\hspace{0.03\textwidth}$ \\
\hline
$a^+_{vB}$      & 1        & $\frac56$ & $\frac12$ & $\frac12$ \\
$a^-_{vB}$      & 0        & $0$       & $-\frac12$ & $-1$       \\
\hline
$a^+_{\kappa B}$& 0        & $-\frac43$ & $0$                  & $0$   \\
$a^y_{\kappa B}$& 0        & $2$        & $1$                  & $0$   \\
\hline
$b^+_{vB}$      & 0        & $\frac13$  & $-1$                 & $0$   \\
$b^-_{vB}$      & 0        & $0$        & $1$                  & $0$   \\
\hline
$b^+_{\kappa B}$& $1$      & $\frac43$  & $0$                  & $1$   \\
$b^-_{\kappa B}$& $2$      & 0          & $0$                  & $-2$  \\
$b^y_{\kappa B}$& 0        & $-\frac23$ & $2$                  & $0$   \\
\hline
$p^+_{vB}$      &$1 $      & $\frac{11}{24}$ & $\frac{7}{8}$ &$\frac{1}{8}$ \\
$p^-_{vB}$      &$-\frac32$& $0$             & $-\frac{7}{8}$ &$-\frac{1}{4}$\\
\hline
\end{tabular}
\caption{Coefficients $a^\pm_{vB}$, $b^\pm_{vB}$, $a^+_{\kappa B}$, $b^\pm_{\kappa B}$, $a^{y}_{\kappa B}$, $b^{y}_{\kappa B}$ and $p^\pm_{vB}$.}.
\label{tab:abp2}
\end{table}

In the symmetric nuclear matter, all $\beta$-dependent terms vanish, and the approximate solutions~(\ref{eq:approx_ms_sv}) can be written in a simple form:
\begin{eqnarray}
\bar m^*_B = \frac{\bar m_B +[\bar B_B^\kappa \kappa (b^{+}_{\kappa B} + b^{y}_{\kappa B} y) + m_s \bar B_B^v b^{+}_{vB} v]\rho}{
1+[
\bar A_B^g g + \bar A_B^v a^{+}_{vB}
 + m_s \bar A_B^\kappa  \kappa (a^{+}_{\kappa B} + a^{y}_{\kappa B} y)
 ]\rho
 },\ \,
\label{eq:Ms_sim}
 \\
\bar \Sigma^V_B = \frac{-\bar P_B^v p^{+}_{vB} v \rho}{
1+[\bar A_B^g g + \bar A_B^v a^{+}_{vB} + m_s \bar A_B^\kappa  \kappa (a^{+}_{\kappa B} + a^{y}_{\kappa B} y) ]\rho
 }.\ \,
 \label{eq:Sv_sim}
 \end{eqnarray}

\begin{figure}[htb]
\centering
 \includegraphics[width=8cm]{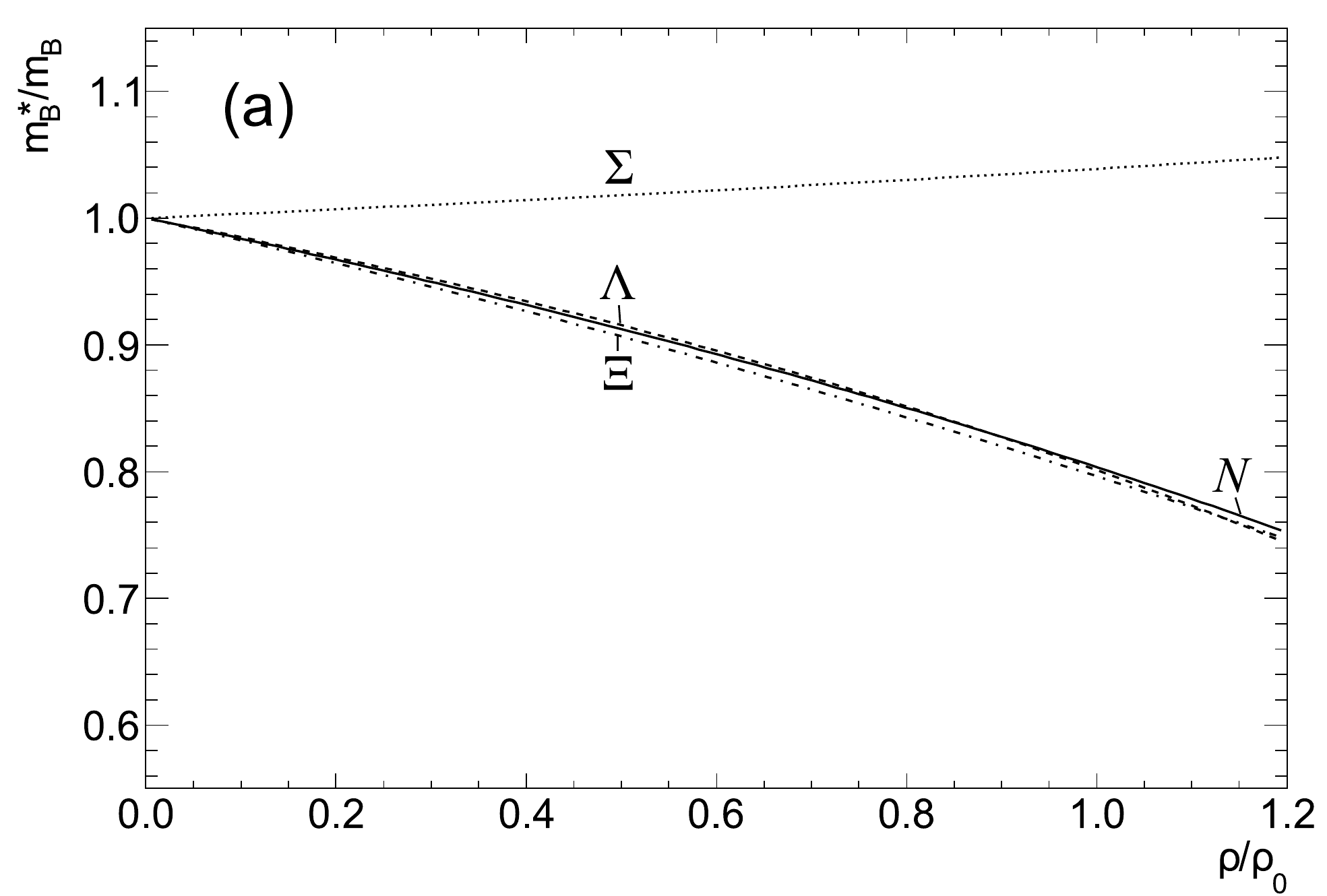}
 \includegraphics[width=8cm]{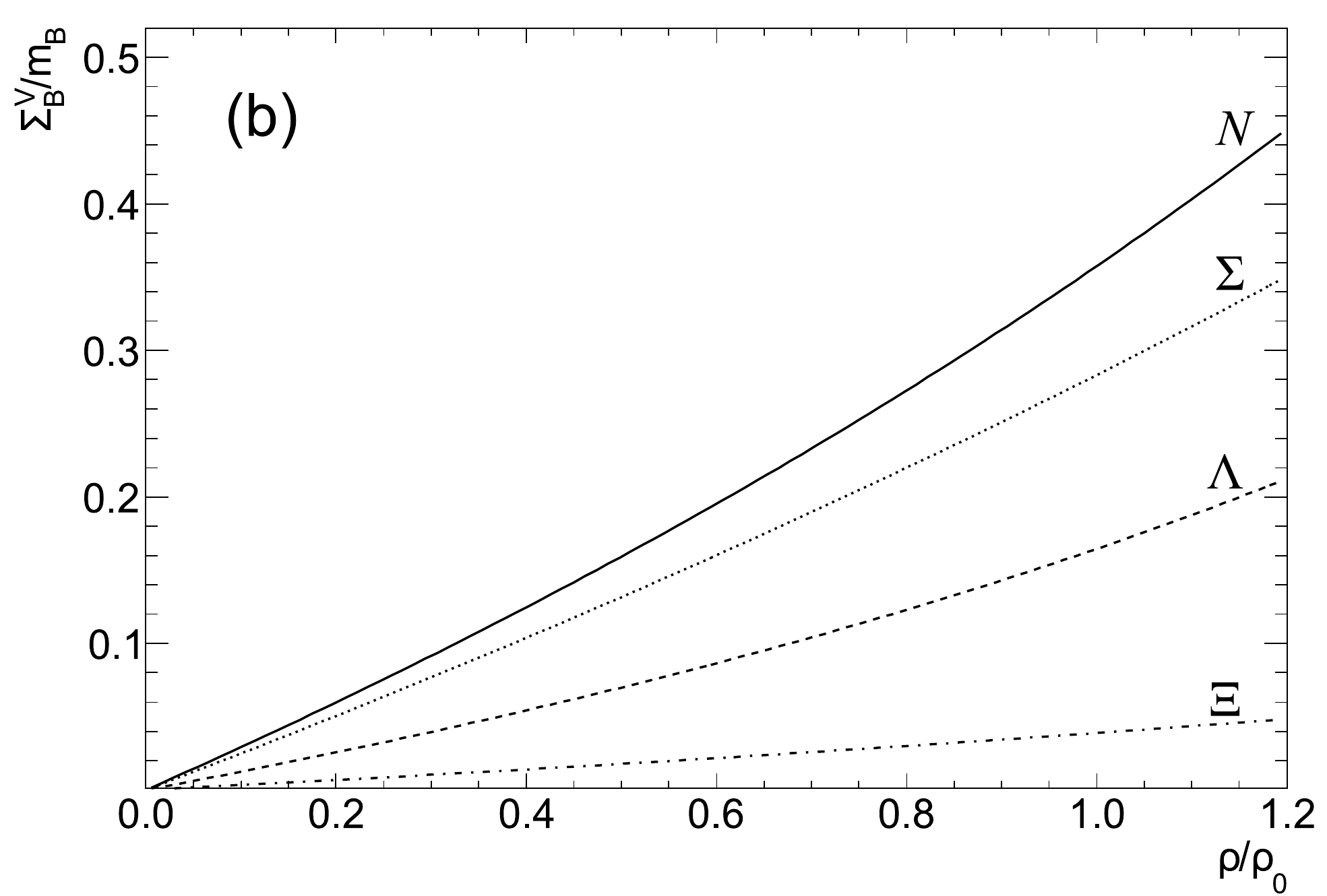}
 \caption{The density dependence of the baryon effective mass $m_B^*$ (a) and the vector self-energy $\Sigma_B^V$ (b) in the symmetric nuclear matter.}
 \label{fig:mssv}
\end{figure}

The obtained density dependence for the baryon effective masses and the vector self-energies is shown in Fig.~\ref{fig:mssv} for the default values $\kappa=8$ and $y=0.08$. One can observe that the ratios $m_B^*/m_B$ for $N$, $\Lambda$ and $\Xi$ follow almost identical trend and are around 0.8 at the saturation density. This coincidence is related to the fact that the main contribution to the effective masses comes from the term proportional to the coefficient $b^+_{\kappa B}$ which equals to 1 for $N$ and $\Xi$ and $\frac43$ for $\Lambda$ (see Table~\ref{tab:abp2}). In case of $\Lambda$, there are two additional terms in the numerator of equation~(\ref{eq:Ms_sim}), proportional to $b^y_{\kappa \Lambda}$ and $b^+_{v \Lambda}$, which compensate the difference of $b^+_{\kappa \Lambda}=\frac43$ from $b^+_{\kappa N}=b^+_{\kappa \Xi}=1$. On the other hand, the coefficient $b^+_{\kappa B}$ for $\Sigma$ is equal to 0 and the main contribution to the density dependence of the $\Sigma$ effective mass comes from the term proportional to the strange quark content $y$ and the strange quark mass $m_s$. The interplay of these terms results in a positive slope of the $\Sigma$ effective mass.

The slope of the vector self-energy $\Sigma_B^V$, shown in Fig.~\ref{fig:mssv}, is basically determined by the coefficient $p^+_{vB}$ in the numerator of equation~(\ref{eq:Sv_sim}). According to Table~\ref{tab:abp2}, the vector self-energies for $N$, $\Lambda$, $\Sigma$ and $\Xi$ should approximately scale as $1:\frac{11}{24}:\frac{7}{8}:\frac{1}{8}$ in contract to the prediction
$1:\frac{2}{3}:\frac{2}{3}:\frac{1}{3}$ of the naive quark model. Different continuum thresholds result in different average values for the functions $\bar P_v^B$ which effects the deviation from the scaling $1:\frac{11}{24}:\frac{7}{8}:\frac{1}{8}$. The ratio of the nucleon vector self-energy to the vacuum nucleon mass is about $36\%$ at saturation density  which is in agreement with results obtained in~\cite{Drukarev2004} for the lowest-order condensates.

It is also instructive to study the dependence of the baryon effective masses and vector self-energies on the expectation value $\kappa$ and the strange quark content $y$ which appear to be somewhat ambiguous according to the discussion of section~\ref{sec:condensates}. As for the vector self-energy, its dependence on the scalar quark expectation values is marginal according to equation~(\ref{eq:Sv_sim}), since it comes only from the term proportional to $m_s$ in the denominator. Moreover, in case of $N$ and $\Xi$ baryons, this dependence completely vanishes since $a^+_{\kappa N} = a^+_{\kappa \Xi} = a^y_{\kappa N} = a^y_{\kappa \Xi} = 0$. Numerical analysis shows that any reasonable variations of $\kappa$ and $y$ do not change the values of the $\Lambda$ and $\Sigma$ vector self-energies within the accuracy of 0.5\%.

\begin{figure}[htb]
\centering
\includegraphics[width=8cm]{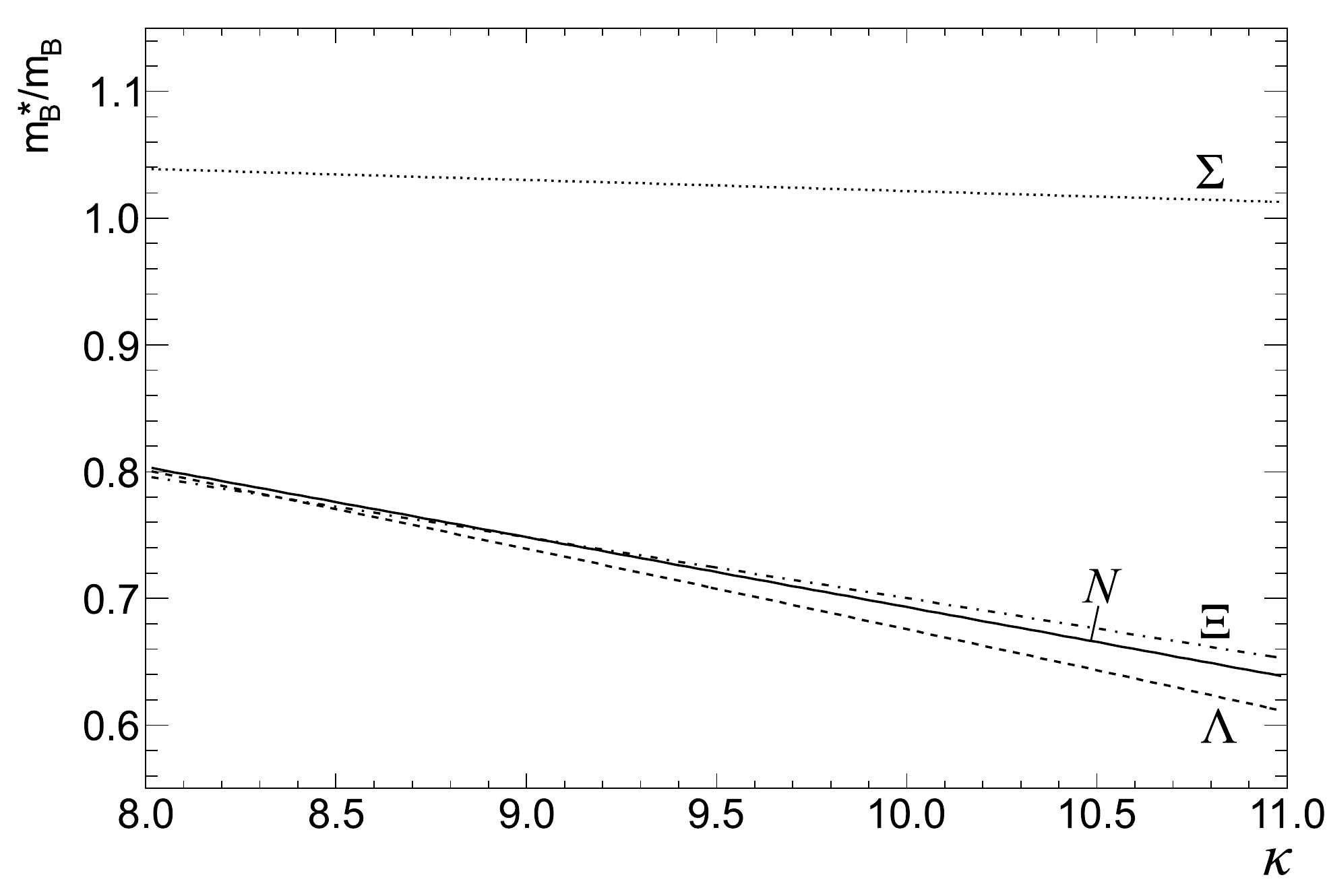}
\caption{Baryon effective masses as functions of $\kappa$ at the saturation density.}
\label{fig:k}
\end{figure}

On the other hand, the parameters  $\kappa$ and $y$ play important roles in the calculation of the baryon effective mass since they appear in the leading terms in the numerator of equation~(\ref{eq:Ms_sim}). Variation of the baryon effective masses versus $\kappa$ at the saturation density is illustrated in Fig.~\ref{fig:k}. One can observe that the effective masses for $N$, $\Lambda$ and $\Xi$ baryons drop down from $0.8 m_B$ to $\sim 0.65 m_B$ when $\kappa$ is varied from the conventional value $\kappa=8$ to the value $\kappa=11$, which is favoured by the recent results on the nucleon $\sigma$ term. On the other hand, $\Sigma$ effective mass has only marginal dependence on $\kappa$ since $b^+_{\kappa \Sigma}=0$.

\begin{figure}[htb]
\centering
\includegraphics[width=8cm]{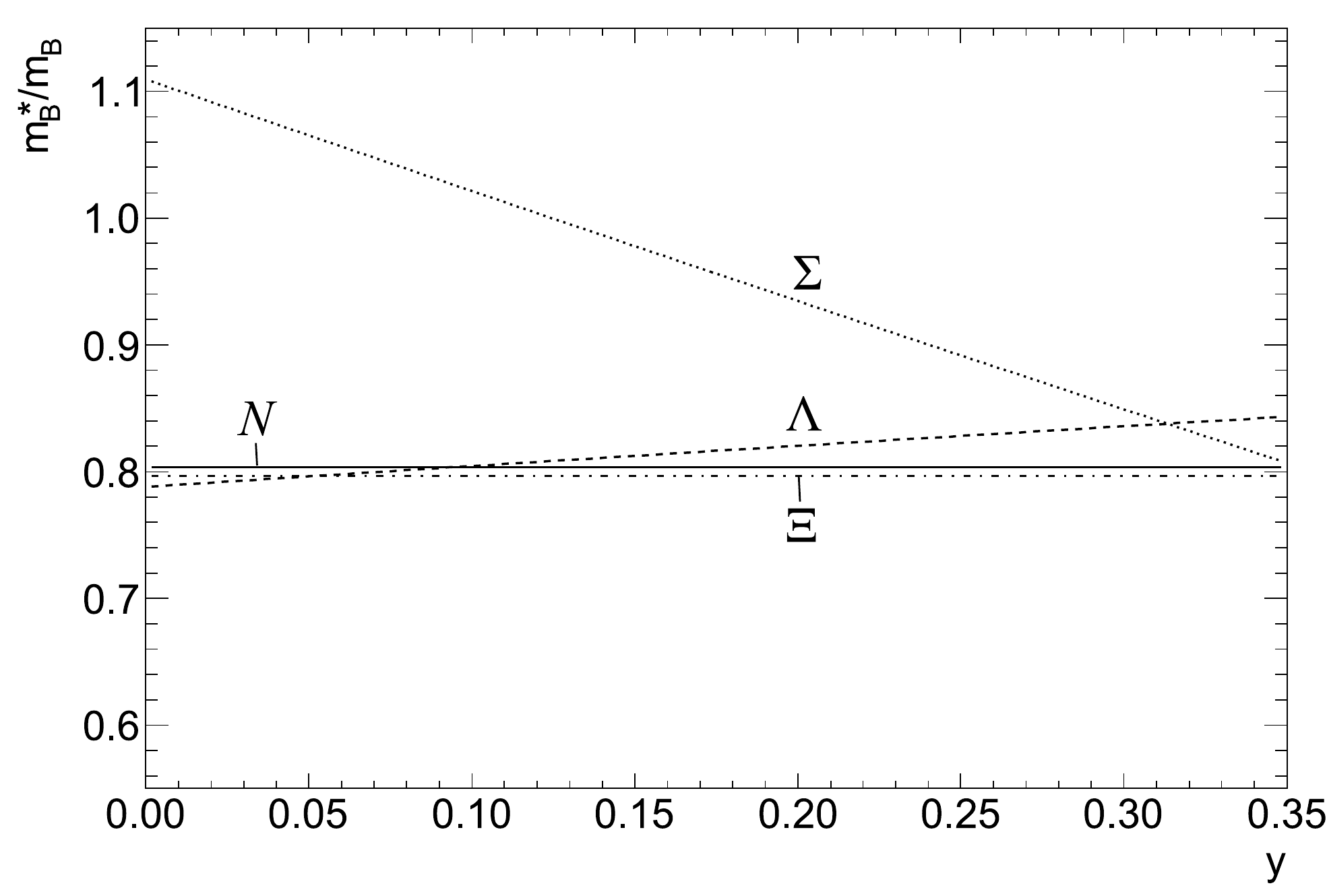}
\caption{Baryon effective masses as functions of $y$ at the saturation density.}
\label{fig:y}
\end{figure}

Finally, the variation of the baryon effective masses versus the strange quark content $y$ is illustrated in Fig.~\ref{fig:y}. In case of $N$ and $\Xi$ baryons, there is no dependence on $y$ since $a^y_{\kappa B} = b^y_{\kappa B} = 0$ for $B=N,\Xi$. The effective mass for the $\Lambda$ hyperon changes by about $5\%$ when the strange quark content $y$ varies from 0 to the somewhat extreme value of 0.35. On the other hand, the effective mass for the $\Sigma$ hyperon dramatically depends on $y$ since the strange scalar quark expectation value appears in the leading term due to vanishing contributions of the light scalar quark expectation values. Numerically, when $y$ increases from 0 to 0.35, the ratio $m^*_\Sigma/m_\Sigma$ drops down from $1.1$ to $0.8$ approaching corresponding values for other baryons. Thus, not only the value but also the sign of the scalar self-energy for the $\Sigma$ hyperon is sensitive to the strange quark content $y$.

\subsection{Asymmetric nuclear matter}
\begin{figure}
\centering
 \includegraphics[width=8cm]{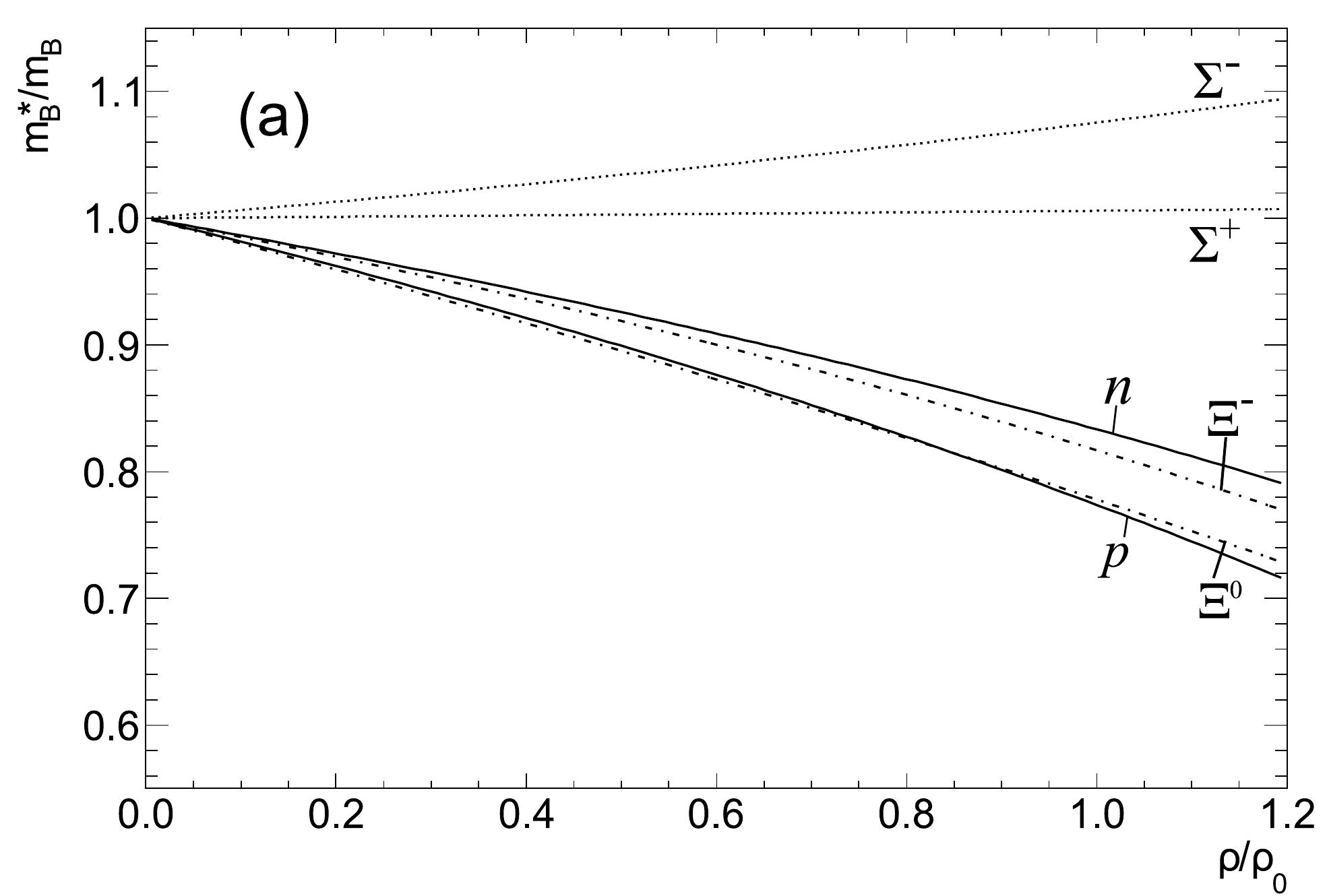}
 \includegraphics[width=8cm]{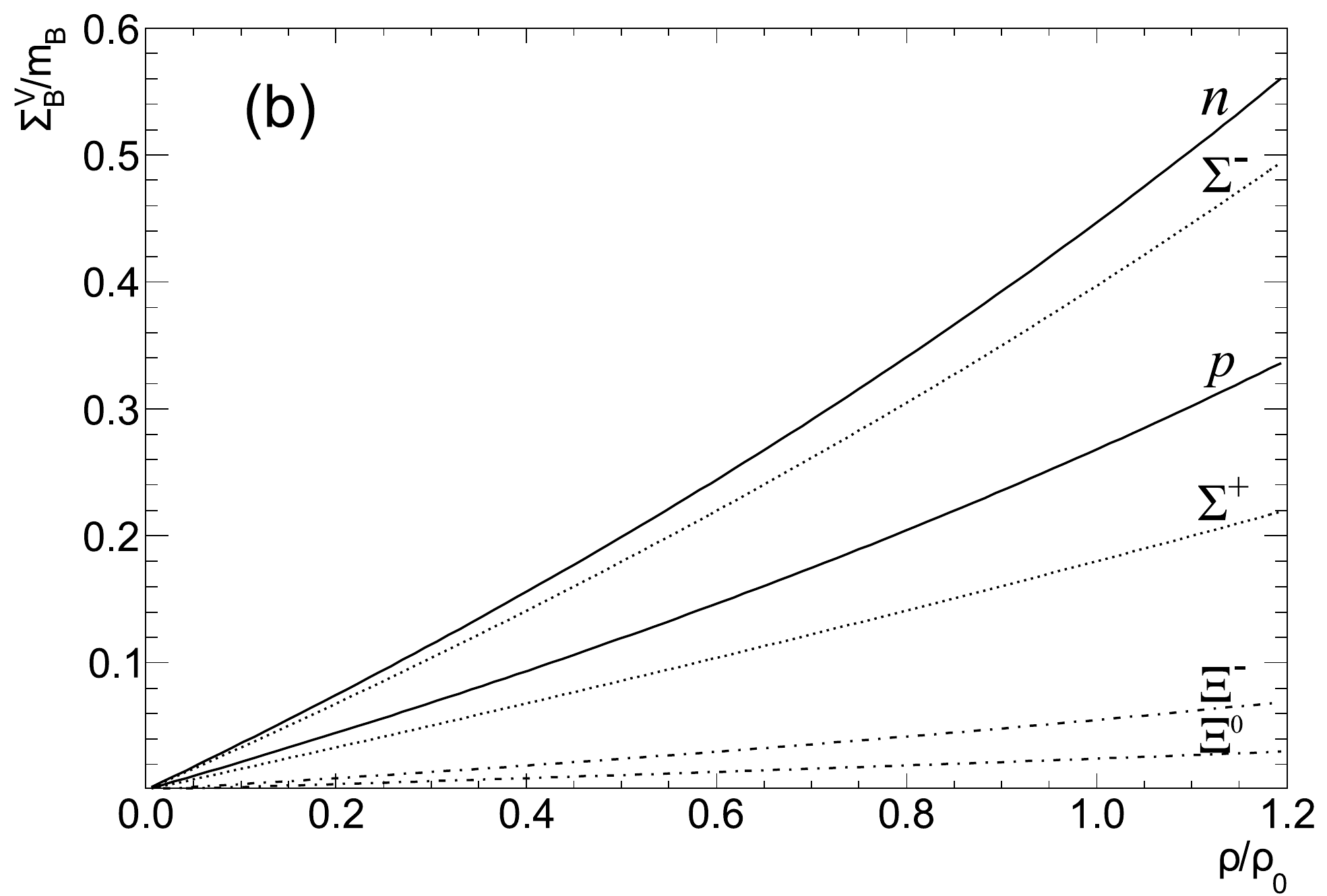}
 \caption{The effective masses (a) and vector self-energies (b) for $p$, $n$, $\Sigma^+$, $\Sigma^-$, $\Xi^0$ and $\Xi^-$ at $\beta=1$ as functions of the baryon density. The calculation was performed at $\kappa=8$ and $y=0.08$.}
 \label{fig:assym}
\end{figure}

Before discussing the asymmetric nuclear matter effects, let us recall general isospin symmetry relations which are automatically satisfied in the sum rule approach:
\begin{eqnarray}
\Sigma^V_p(\beta)&=\Sigma^V_n(-\beta), \qquad\quad m^*_p(\beta)&=m^*_n(-\beta),
\label{eq:isospin_sym1}
\\
\Sigma^V_{\Sigma^+}(\beta)&=\Sigma^V_{\Sigma^-}(-\beta)
, \qquad m^*_{\Sigma^+}(\beta)&=m^*_{\Sigma^-}(-\beta),
\label{eq:isospin_sym2}
\\
\Sigma^V_{\Xi^0}(\beta)&=\Sigma^V_{\Xi^-}(-\beta)
, \qquad m^*_{\Xi^0}(\beta)&=m^*_{\Xi^-}(-\beta),\qquad \quad
\label{eq:isospin_sym3}
\end{eqnarray}
while $\Lambda$ and $\Sigma^0$ self-energies do not depend on the isospin asymmetry. In the following, we  will consider the extreme case of neutron matter with the asymmetry parameter $\beta=1$ for $p$, $n$, $\Sigma^+$, $\Sigma^-$, $\Xi^0$ and $\Xi^-$, while the effects of another extreme case of the proton matter with $\beta=-1$ can be obtained from the isospin symmetry relations~(\ref{eq:isospin_sym1}--\ref{eq:isospin_sym3}).
The effective masses and vector self-energies for $p$, $n$, $\Sigma^+$, $\Sigma^-$, $\Xi^0$ and $\Xi^-$ at $\beta=1$ are shown in Fig.~\ref{fig:assym} as functions of the baryon density.

The splitting of the effective masses for nucleons is determined by the $I$ structure~(\ref{eq:XBI}) resulting in negative proton-neutron mass difference $\Delta m_{pn}^*=m_{p}^*-m_{n}^*<0$ in the neutron matter. This result is in agreement with leading OPE calculations in~\cite{Drukarev2004}. Note, however, that inclusion of higher-order OPE contributions would provide $\Delta m_{pn}^*>0$ which is expected in the relativistic approaches~\cite{Drukarev2004}. In contrast, the dominant contribution to the effective mass splitting for $\Sigma$ and $\Xi$ hyperons comes from the $q$ structure~(\ref{eq:XBq}) resulting in relations $m^*_{\Sigma^-}>m^*_{\Sigma^0}>m^*_{\Sigma^+}$ and $m^*_{\Xi^{-}}>m^*_{\Xi^0}$ in the neutron matter. Numerically, the splitting at the saturation density is of about $6\%$, $7\%$ and $4\%$ of the corresponding mass for $N$, $\Sigma$ and $\Xi$ baryons.

The splitting of the vector self-energies is dominated by the contribution of the $p$ structure~(\ref{eq:XBp}). The relative strength of the splitting for different baryons mainly comes from the product $p^-_{vB}I_{3B}$, thus $N$, $\Sigma$ and $\Xi$ splittings should approximately scale as $6:7:1$ in contrast to the naive quark counting model which predicts the relation $1:2:1$. Numerically, the vector self-energy splitting at the saturation density is of about 170, 260 and 40 MeV  for $N$, $\Sigma$ and $\Xi$ baryons respectively in agreement with the approximate relation $6:7:1$.

\section{Discussion}
\label{sec:dis}
In this section, we compare the obtained results with experimental data and alternative theoretical models. The nucleon in-medium properties have been already studied in detail in~\cite{Drukarev2004a,Drukarev2004}, therefore we will concentrate on the hyperon case. We will discuss the solutions of the sum rule equations obtained in the approximation of the vacuum continuum thresholds~(\ref{eq:approx_ms_sv}) keeping in mind that exact solutions based on $\chi^2$ fit~(\ref{eq:chi2}) may lead to somewhat different results.

Let us start with the $\Lambda$ hyperon. The obtained scalar self-energy $\Sigma^S_\Lambda(\rho_0) \sim -210$ MeV  is in good agreement with the BHF calculations~\cite{Schulze1998} where the value $m^*_\Lambda/m_\Lambda = 0.84$ was reported. Besides, the ChPT approach~\cite{Korpa2001} provided the scalar self-energy about 55 MeV at $0.4 \rho_0$, which is close to the value 65 MeV obtained with the sum rule approach. As for the vector self-energy, the value $\Sigma_\Lambda^V(\rho_0)  = 180$ MeV was obtained providing a non-relativistic potential $U_\Lambda(\rho_0) \sim - 30$ MeV which is in surprisingly perfect agreement with the hypernuclear data~\cite{Millener1988} and the BHF calculations~\cite{Schulze1998}. Recall, however, that the $\Lambda$ effective mass is highly sensitive to $\kappa =\langle N|\bar uu+\bar dd|N \rangle$ and hence to the value of the $\sigma_{\pi N}$ term. Changing the value from $\sigma_{\pi N} = 45$ MeV to 60 MeV would give much higher scalar self-energy and the potential would appear to be much deeper.

On the other hand, the value of the effective mass $m^*_\Sigma$ dramatically depends on the strange quark content $y$. Assuming $y<0.12$, the value of $m^*_\Sigma$ would grow with density in agreement with the ChPT predictions~\cite{Korpa2001}. However, for the larger strange quark content values, one would get the $\Sigma$ effective mass decreasing with density in support of the BHF approach~\cite{Schulze1998}. In any case, account of the large vector self-energy $\Sigma^V_\Sigma (\rho_0) \approx 0.3 m_\Sigma$ results in the repulsive $\Sigma$ hyperon potential at the saturation density in agreement with experimental data~\cite{Bart1999,Batty1994,Batty1994a,Mares1995}. Touching the $\Sigma$ isospin triplet in the neutron matter, we obtain the ordering $\Sigma^+$, $\Sigma^0$, $\Sigma^-$ in increasing mass shift similar to BHF calculations~\cite{Baldo1998}. Note, also, that the ChPT calculations~\cite{Korpa2001} for the neutron matter provided anomalous ordering of the effective masses $m^*_{\Sigma^0} > m^*_{\Sigma^-}$ due to large isospin-symmetry violation effects.

As for the $\Xi$ hyperon, the density dependence of the scalar self-energy appears to be similar to the nucleon case in contrast to the expectations from the naive quark model $\Sigma^S_\Xi/\Sigma^S_N = 1/3$. On the other hand, the $\Xi$ vector self-energy accounts about $1/8$ of the nucleon self-energy. The large scalar self-energy in combination with the small $\Sigma_\Xi^V$ provides the attractive potential $U(\rho_0)$ about $-200$ MeV which is an order of magnitude larger than the value $-18$ MeV extracted from the experimental data~\cite{Dover1983,Fukuda1998,Khaustov2000}. Unfortunately, the $\Xi$ hyperon is not considered in ChPT or BHF approaches due to lack of $\Xi N$ scattering data. Note, however, that the Gell-Mann--Okubo--like formulas~(\ref{eq:GMOms}--\ref{eq:GMOsv}) could serve as an $SU(3)$ motivated way for the tuning of the $\Xi$--hyperon in-medium potential.

The hyperon in-medium properties were also studied in the RMF framework~\cite{Glendenning1991,Schaffner1996} where meson-hyperon coupling constants were fixed to reproduce the hyperon in-medium potentials at saturation density. This approach suffers from large ambiguities since the isoscalar $\sigma$ and $\omega$ meson couplings appear to be highly correlated while the isovector meson couplings remain unconstrained by the hypernuclear data.
As it was discussed in~\cite{Drukarev2004a}, the lowest-order OPE terms, considered in this paper, correspond to exchanges by localized quark-antiquark pairs or effective vector and scalar mesons between baryons and nucleons in nuclear matter. Therefore one could use the obtained results as an input for the calculation of the effective meson-baryon coupling constants in the RMF approach. 

However, higher-order OPE terms appear to be numerically important. For example, inclusion of nonlocal vector condensate and four-quark contributions would subtract 60 MeV and 110 MeV from the lowest dimension value $\Sigma^V_N(\rho_0) = 270$ MeV~\cite{Drukarev2004a}. The scalar self-energy $\Sigma^S_N(\rho_0)=-140$ MeV would remain almost unchanged because the four-quark condensates and nonlocal contributions would add about $-100$ MeV and $100$ MeV, respectively~\cite{Drukarev2004a}. Similarly, the higher-order OPE terms should play an important role in the hyperon sum rules.

\section{Summary}
\label{sec:sum}

The QCD sum rules provide a unique consistent formalism for investigation of the baryon octet in-medium properties. In contrast to other approaches, the QCD sum rules do not rely on phenomenological parameters of the baryon-meson interactions.

In this paper, the baryon effective masses and vector self-energies were expressed in terms of a few in-medium QCD condensates of the lowest dimension which have been either calculated or related to the observables. It was shown, that the effective masses and vector self-energies in the baryon octet obey the relations similar to the Gell-Mann--Okubo mass formulas up to the linear $SU(3)$-breaking terms. Moreover, the coefficients in the OPE terms provide a peculiar $SU(3)$-breaking pattern, e. g. vector self-energies in the symmetric nuclear matter are predicted to scale approximately as $1:\frac{11}{24}:\frac{7}{8}:\frac{1}{8}$ for the $N$, $\Lambda$, $\Sigma$ and $\Xi$ baryons, respectively.

Numerical studies for the $N$, $\Lambda$, $\Sigma$ and $\Xi$ baryon properties were carried out both in the symmetric and asymmetric nuclear matter in the approximation that in-medium effective continuum thresholds do not depend on density and remain equal to vacuum ones. The hyperon effective masses reveal a strong dependence on the values of $\sigma_{\pi N}$ term and the strange quark content $y$ which are  known with poor accuracy. Nevertheless, the obtained effective masses and vector self-energies are in reasonable agreement with the results from other nuclear physics methods.

The provided formalism can be extended to the case of matter composed of an arbitrary mixture of baryons, which is important in the calculations of the neutron star equation of state. Besides, contributions of the higher-dimensional condensates and radiative corrections could be included in order to improve the accuracy of the method.

\begin{acknowledgments}
I am grateful to E. G. Drukarev, V. A. Sadovnikova and M. G. Ryskin for many fruitful discussions on the subject.
\end{acknowledgments}

\bibliography{ref}

\end{document}